\documentclass[showpacs,prl,a4paper,twocolumn]{revtex4}
\usepackage{epsfig}

\begin{document}
\newcommand{\degree}{$^{\rm\circ} $}
\newcommand{\pcite}{\protect\cite}
\newcommand{\pref}{\protect\ref}

\title{DNA Dynamics in A Water Drop}

\author{Alexey K. \surname{Mazur}}
\email{alexey@ibpc.fr}
\altaffiliation{FAX:+33[0]1.58.41.50.26}
\affiliation{Laboratoire de Biochimie Th\'eorique, CNRS UPR9080,
Institut de Biologie Physico-Chimique,
13, rue Pierre et Marie Curie, Paris,75005, France}
 
\begin{abstract}
Due to its polyionic character the DNA double helix is stable and
biologically active only in salty aqueous media where its charge
is compensated by solvent counterions. Monovalent metal ions are
ubiquitous in DNA environment and they are usually considered as the
possible driving force of sequence-dependent modulations of DNA
structure that make it recognizable by proteins. In an effort to
directly examine this hypothesis, MD simulations of DNA in a water
drop surrounded by vacuum were carried out, which relieves the
requirement of charge neutrality. Surprisingly, with zero
concentration of counterions a dodecamer DNA duplex appears metastable
and its structure remains similar to that observed in experiments.
\end{abstract}


\maketitle

\ \\

It has been long recognized that, because of the polyionic nature of
DNA, solvent counterions are required for its stability, and that
gross structural changes in DNA can be provoked by changing the
concentration, the charge or the type of counterions
\cite{Saenger:84}. More recently it has been proposed that free ions
might also act as biological regulators because, by binding to DNA,
they can provoke conformational deformations recognized by specific
proteins
\cite{Jayaram:96,Young:97a,Young:97b,Hud:97,Rouzina:98,Williams:00}.
The most controversial is the role of the common monovalent cations
Na$^+$ and K$^+$. They are ubiquitous in the DNA environment and can
be readily available for any purpose. Until recently, they remained
invisible in experimental DNA structures because it is difficult to
detect them in water, and it has been suggested that they are perhaps
responsible for the most widespread deformations of the double helix,
namely, narrowing of the minor groove and bending. It is assumed that
counterions are sequestered in the minor groove of some sequences,
which brakes the symmetry of the repulsive electrostatic forces and
provokes deformations. This model is general and it easily explains
other puzzling effects in DNA structure.

The above hypothesis is supported by many recent studies. Penetration
of monovalent cations into the minor DNA groove has been confirmed by
X-ray diffraction
\cite{Rosenberg:73,Bartenev:83,Tereshko:99b,Sines:00},
NMR spectroscopy \cite{Hud:99,Denisov:00}, and MD simulations
\cite{Young:97a,Feig:99,Strahs:00,Stefl:00,Hamelberg:01}.
However, it appears difficult to find a discriminating set-up for
testing the cause and consequence relationship between the ions and
the DNA structure. All available evidences have more than one
interpretation making this problem highly controversial
\cite{McFail-Isom:99,Chiu:99,McConnell:00}. For example, correlations
between ion positions and the local groove width observed in MD
cannot answer whether the ions perturb DNA or they just bind
``opportunistically'' in the sites of low potential near already
deformed double helix \cite{Hamelberg:01}. To clarify the
issue of cause and effect one would have to remove solvent ions and
check if the supposed counterion effects disappear with them.
Unfortunately, a counterion-free DNA does not exist in nature whereas
the most reliable computational procedures presently employed require
that the simulation cell that holds DNA carries zero net charge.
Therefore, in both simulations and experiments the counterion effects
cannot be completely eliminated.

In an effort to directly address this question I have adapted the
Particle Mesh Ewald algorithm \cite{Darden:93,Essmann:95} for modeling
dynamics of DNA in a water drop surrounded by vacuum. Particle-mesh
calculations with vacuum boundary conditions are long known in physics
\cite{Hockney:81}, but, to my knowledge, they were never applied
to chemical or biological systems. Free vacuum boundaries are
intuitively most simple and they allow one to relieve the problems of
charge neutrality and possible artifacts from interactions between
periodical images. I describe here the first such ``naive'' all atom
simulations of DNA in water, with unperturbed Coulomb electrostatics.
It appears that, with zero concentration of counterions, a dodecamer
DNA duplex is metastable and its structure remains similar to that
observed in experiments.

The Dickerson-Drew dodecamer (CGCGAATTCGCG, \cite{Wing:80}) in a
canonical B-DNA conformation \cite{Arnott:72} is surrounded by a
spherical drop of 4000 TIP3P water molecules \cite{Jorgensen:83}.
Initially, the drop had around 50 \AA\ in diameter and in dynamics it
remained roughly spherical. A rectangular unit cell is constructed
around the drop with the minimal separation of 25 \AA\ between the
water molecules and the cell sides.  The cell is replicated
periodically, which gives an infinite lattice of water drops with at
least  50 \AA\ spacing between the closest neighbors. A shifted
Coulomb's law is used, with $U(r_{ij})=z_iz_j(1/r_{ij}-1/R_0)$ for
$r_{ij}<R_0$ and $U(r_{ij})=0$ for longer distances, where $R_0=50$
\AA, which eliminates any interactions between periodical images.
Because this shifting does not affect the forces, the system behaves
in dynamics as if surrounded by infinite vacuum.  Within the drop the
electrostatics are effectively evaluated with a cut-off of 50 \AA,
which is larger than the DNA size and applies only to a small fraction
water molecules at opposite poles of the drop. Larger distances were
also tested, but showed no noticeable difference. The Van-der-Waals
interactions are computed with a conventional cut-off of 9 \AA.

\begin{table}[t]\caption{\label{Tenpa} Some structural
parameters
of standard and computed DNA conformations. Sequence averaged
helical parameters were computed with program Curves
\pcite{Curves:}.
All distances are in angstr{\"o}ms and angles in degrees. }
\begin{ruledtabular}
\begin{tabular}[t]{|ccccccc|}
           & Xdisp
             & Inclin
               & Rise
                 & Twist
                   & RMSD-A\footnotemark[1]
                     & RMSD-B\footnotemark[1]\\
\hline
A-DNA & -5.4 & +19.1 & 2.6 & 32.7 & 0.0  & 6.2  \\
B-DNA & -0.7 & -6.0  & 3.4 & 36.0 & 6.2  & 0.0  \\
\hline
Tj1\footnotemark[2]
      & -2.2 & +6.9  & 3.3 & 33.8 & 4.52 & 2.36 \\
Tj2\footnotemark[3]
      & -2.0 & +6.5  & 3.2 & 34.1 & 4.57 & 2.18 \\
Tj3\footnotemark[4]
      & -2.1 & +6.9  & 3.3 & 33.6 & 4.46 & 2.41 \\
\end{tabular}
\end{ruledtabular}
\footnotetext[1]{Heavy atom root mean square deviation from
the corresponding canonical DNA form.}
\footnotetext[2]{Water drop calculation without counterions.}
\footnotetext[3]{Water drop neutralized by 22 $\rm Na^+$ ions.}
\footnotetext[4]{Conventional PME calculation with
counterions and periodical boundaries.}
\end{table}

\begin{figure}
\centerline{\psfig{figure=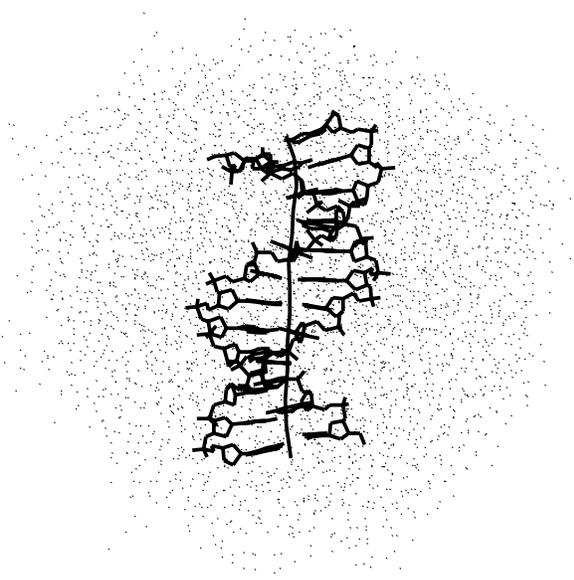,width=9cm,angle=0.}}
\caption{\label{Fsnap}
A snapshot from the last nanosecond of the trajectory with
water oxygen positions shown by dots.
}\end{figure}

The shifted Coulomb interactions are evaluated by using the SPME
method \cite{Essmann:95}. The shifting is taken into account in the
reciprocal sum by replacing the Fourier transform of $1/r$ with that
of the shifted law. The resulting series is absolutely convergent
regardless of the system charge. The values of energies and forces
thus obtained have been verified by direct computation without
cut-offs. Modifications in the direct sum are not necessary if we are
only interested in the forces and not in the absolute energy values.
It can be shown that this simple procedure is essentially equivalent to
mathematically more complex earlier formulations of the Ewald method for
cluster calculations \cite{Hockney:81,Martyna:99}.
In dynamics, the shape of the drop fluctuates and the size of the unit
cell is adjusted accordingly at every time step. Evaporated water
molecules are detected and excluded from calculations to prevent
explosion in the cell size. After every 50 ps the calculation is
stopped and water molecules that have left the drop are re-introduced
with zero velocities by scattering them randomly near the surface of
the drop. The rate of evaporation was around 84 mol/ns, that is on
average four molecules had to be re-introduced at each stop.

All calculations were carried out with AMBER98 parameters
\cite{Cornell:95,Cheatham:99} by using the ICMD method
\cite{Mzjcc:97,Mzjchp:99,Mzbook:01} with the time step of 0.01 ps. Two
control simulations of the same dodecamer have been carried out for
comparison.  In the first, 22 $\rm Na^+$ ions were included in the
same water drop. The second represents a conventional calculation in a
rectangular unit cell including DNA, 3901 water molecules, and 22 $\rm
Na^+$ ions by the original SPME method with periodical boundary
conditions. The trajectories were continued to 5 ns.

Figure \ref{Fsnap} shows a snapshot of the charged drop from the last
nanosecond of the trajectory. The system appears metastable in the
nanosecond time range. The DNA molecule shows no signs of
denaturation. The water media remains continuous without internal
bubbles. The surface of the drop is covered by spiky ``protuberances''
which reduce with increased drop size and disappear if DNA is
neutralized by counterions. The DNA structures averaged over the last
2.5 ns of the three trajectories are characterized in Table
\ref{Tenpa}. These data indicate that with zero counterion
concentration this dodecamer DNA molecule remains in B-form. It is
understood that with increased DNA length it should have exploded
because otherwise the electrostatic energy would go to infinity.
However, no such trend is seen in Table \ref{Tenpa}, which
indicates that critical lengths of catastrophic deformations are much
larger whereas the dodecamer B-form is metastable in water even if it
is charged.

\begin{figure}
\centerline{\psfig{figure=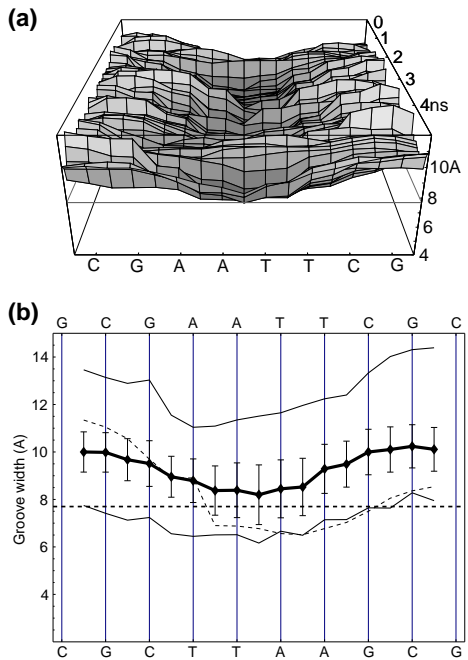,width=7cm,angle=0.}}
\caption{\label{Fmg}
(a) The time evolution of the profile of the minor groove in Tj1. The
surface plots are formed by time-averaged successive minor groove
profiles, with that on the front face corresponding to the final DNA
conformation. The groove width is evaluated by using space traces of
C5' atoms \pcite{Mzjmb:99}. Its value is given in angstr\"oms and the
corresponding canonical B-DNA level of 7.7 \AA\ is marked by the thin
straight lines on the faces of the box. Note that the groove width can
be measured only starting from the third base pair from both
termini.\\
(b) The minor groove profile averaged over the last nanosecond of Tj1.
The central trace represents the average groove width with rms
fluctuations shown as error bars. The upper and lower solid traces
show the maximal and minimal values, respectively. The dotted trace
exhibits the profile observed in the experimental X-ray structure
\pcite{Wing:80}. The canonical B-DNA groove width is marked by the
horizontal dashed line. Despite the narrowing, the groove width
remains larger than the canonical value \pcite{Mzjmb:99}, which
corresponds to the lower average twist. }\end{figure}

The scales of differences between the DNA conformations in the three
trajectories are well characterized by the data in Table \ref{Tenpa}
and they were always far from statistically significant. The
corresponding experimental data are available only for the twist,
namely, its value drops by 0.1\degree\ when the NaCl concentration is
reduced from 0.3M to 0.05M \cite{Anderson:78}. A shift by 0.1\degree\
in the average twist is too small to be detected in a 5 ns simulation
because, for this relatively small molecule, its fluctuations between
consecutive 1 ns averaged structures can reach 1.0\degree. The
absolute twist value is lower than in experiment, which is a known
general feature earlier discussed in the literature
\cite{Cheatham:99}.

The most famous feature of this DNA molecule is the middle AATT
fragment. It is long known from experiments that the minor DNA groove
always narrows in this and some similar sequences, called A-tracts,
and widens outside of them. Figure \ref{Fmg} exhibits dynamics of
the minor groove in Tj1 and its average profile over the last
nanosecond.  It has a characteristic waving shape with a narrowing in
the middle.  The amplitude of this modulation is similar to that in
the experimental X-ray structure. The minimal width is 1.5 \AA\ larger
than experimental value, which is probably linked mechanically to the
lower average twist.  Similar results were obtained for the other two
trajectories and they are close to earlier reported simulation studies
carried out with non-zero counterion concentrations \cite{Young:97b}.

The sequence-dependent groove-width modulations in DNA are well
established experimentally and, in the recent years, they have been
proposed to result from interactions with bound monovalent metal ions
commonly undetectable in X-ray crystal maps
\cite{Jayaram:96,Williams:00,Hamelberg:01}. The present results
evidence that it is not the case, supporting recent conclusions of
different groups \cite{Denisov:00,McConnell:00,Chiu:99}. They explain
also why groove modulations and intrinsic DNA bending could be
reproduced in MD simulations with simplified treatment of
electrostatic interactions that ignored specific counterion effects
\cite{Mzjacs:98,Mzjacs:00,Mzjcc:01}.

According to the counterion condensation theory, DNA in aqueous
environment should be always covered by a shell of counterions and its
charge should be compensated by around 75\% regardless of the bulk ion
concentrations \cite{Manning:78}. The results presented here do not
contradict this theory but they are somewhat at odds with an implicit
assumption that the counterion cloud is critical for the native DNA
structure. This was surprising, at least for the author, and suggests
that we are still far from complete understanding of interactions that
control the DNA structure. The correlations observed in earlier MD
simulations\cite{Hamelberg:01} apparently are due to
binding of counterions in sites of low potential near an already
narrowed minor groove, therefore, these interactions are
structure-specific rather than sequence-specific, and they cannot be
the driving force of the corresponding DNA deformations.


\begin{thebibliography}{39}
\expandafter\ifx\csname natexlab\endcsname\relax\def\natexlab#1{#1}\fi
\expandafter\ifx\csname bibnamefont\endcsname\relax
  \def\bibnamefont#1{#1}\fi
\expandafter\ifx\csname bibfnamefont\endcsname\relax
  \def\bibfnamefont#1{#1}\fi
\expandafter\ifx\csname citenamefont\endcsname\relax
  \def\citenamefont#1{#1}\fi
\expandafter\ifx\csname url\endcsname\relax
  \def\url#1{\texttt{#1}}\fi
\expandafter\ifx\csname urlprefix\endcsname\relax\def\urlprefix{URL }\fi
\providecommand{\bibinfo}[2]{#2}
\providecommand{\eprint}[2][]{\url{#2}}

\bibitem[{\citenamefont{Saenger}(1984)}]{Saenger:84}
\bibinfo{author}{\bibfnamefont{W.}~\bibnamefont{Saenger}},
  \emph{\bibinfo{title}{Principles of Nucleic Acid Structure}}
  (\bibinfo{publisher}{Springer Verlag}, \bibinfo{address}{New York},
  \bibinfo{year}{1984}).

\bibitem[{\citenamefont{Jayaram and Beveridge}(1996)}]{Jayaram:96}
\bibinfo{author}{\bibfnamefont{B.}~\bibnamefont{Jayaram}} \bibnamefont{and}
  \bibinfo{author}{\bibfnamefont{D.~L.} \bibnamefont{Beveridge}},
  \bibinfo{journal}{Annu. Rev. Biophys. Biomol. Struct.}
  \textbf{\bibinfo{volume}{25}}, \bibinfo{pages}{367} (\bibinfo{year}{1996}).

\bibitem[{\citenamefont{Young et~al.}(1997{\natexlab{a}})\citenamefont{Young,
  Jayaram, and Beveridge}}]{Young:97a}
\bibinfo{author}{\bibfnamefont{M.~A.} \bibnamefont{Young}},
  \bibinfo{author}{\bibfnamefont{B.}~\bibnamefont{Jayaram}}, \bibnamefont{and}
  \bibinfo{author}{\bibfnamefont{D.~L.} \bibnamefont{Beveridge}},
  \bibinfo{journal}{J. Am. Chem. Soc.} \textbf{\bibinfo{volume}{119}},
  \bibinfo{pages}{59} (\bibinfo{year}{1997}{\natexlab{a}}).

\bibitem[{\citenamefont{Williams and Maher}(2000)}]{Williams:00}
\bibinfo{author}{\bibfnamefont{L.~D.} \bibnamefont{Williams}} \bibnamefont{and}
  \bibinfo{author}{\bibfnamefont{L.~J.} \bibnamefont{Maher},
  \bibfnamefont{III}}, \bibinfo{journal}{Annu. Rev. Biophys. Biomol. Struct.}
  \textbf{\bibinfo{volume}{29}}, \bibinfo{pages}{497} (\bibinfo{year}{2000}).

\bibitem[{\citenamefont{Young et~al.}(1997{\natexlab{b}})\citenamefont{Young,
  Ravishanker, and Beveridge}}]{Young:97b}
\bibinfo{author}{\bibfnamefont{M.~A.} \bibnamefont{Young}},
  \bibinfo{author}{\bibfnamefont{G.}~\bibnamefont{Ravishanker}},
  \bibnamefont{and} \bibinfo{author}{\bibfnamefont{D.~L.}
  \bibnamefont{Beveridge}}, \bibinfo{journal}{Biophys. J.}
  \textbf{\bibinfo{volume}{73}}, \bibinfo{pages}{2313}
  (\bibinfo{year}{1997}{\natexlab{b}}).

\bibitem[{\citenamefont{Hud and Feigon}(1997)}]{Hud:97}
\bibinfo{author}{\bibfnamefont{N.~V.} \bibnamefont{Hud}} \bibnamefont{and}
  \bibinfo{author}{\bibfnamefont{J.}~\bibnamefont{Feigon}},
  \bibinfo{journal}{J. Am. Chem. Soc.} \textbf{\bibinfo{volume}{119}},
  \bibinfo{pages}{5756} (\bibinfo{year}{1997}).

\bibitem[{\citenamefont{Rouzina and Bloomfield}(1998)}]{Rouzina:98}
\bibinfo{author}{\bibfnamefont{I.}~\bibnamefont{Rouzina}} \bibnamefont{and}
  \bibinfo{author}{\bibfnamefont{V.~A.} \bibnamefont{Bloomfield}},
  \bibinfo{journal}{Biophys. J.} \textbf{\bibinfo{volume}{74}},
  \bibinfo{pages}{3152} (\bibinfo{year}{1998}).

\bibitem[{\citenamefont{Rosenberg et~al.}(1973)\citenamefont{Rosenberg, Seeman,
  Kim, Suddath, Nicholas, and Rich}}]{Rosenberg:73}
\bibinfo{author}{\bibfnamefont{J.~M.} \bibnamefont{Rosenberg}},
  \bibinfo{author}{\bibfnamefont{N.~C.} \bibnamefont{Seeman}},
  \bibinfo{author}{\bibfnamefont{J.~J.~P.} \bibnamefont{Kim}},
  \bibinfo{author}{\bibfnamefont{F.~L.} \bibnamefont{Suddath}},
  \bibinfo{author}{\bibfnamefont{H.~B.} \bibnamefont{Nicholas}},
  \bibnamefont{and} \bibinfo{author}{\bibfnamefont{A.}~\bibnamefont{Rich}},
  \bibinfo{journal}{Nature} \textbf{\bibinfo{volume}{243}},
  \bibinfo{pages}{150} (\bibinfo{year}{1973}).

\bibitem[{\citenamefont{Bartenev et~al.}(1983)\citenamefont{Bartenev,
  Golovanov, Kapitonova, Mokulskii, Volkova, and Skuratovskii}}]{Bartenev:83}
\bibinfo{author}{\bibfnamefont{V.~N.} \bibnamefont{Bartenev}},
  \bibinfo{author}{\bibfnamefont{E.~I.} \bibnamefont{Golovanov}},
  \bibinfo{author}{\bibfnamefont{K.~A.} \bibnamefont{Kapitonova}},
  \bibinfo{author}{\bibfnamefont{M.~A.} \bibnamefont{Mokulskii}},
  \bibinfo{author}{\bibfnamefont{L.~I.} \bibnamefont{Volkova}},
  \bibnamefont{and} \bibinfo{author}{\bibfnamefont{I.~Y.}
  \bibnamefont{Skuratovskii}}, \bibinfo{journal}{J. Mol. Biol.}
  \textbf{\bibinfo{volume}{169}}, \bibinfo{pages}{217} (\bibinfo{year}{1983}).

\bibitem[{\citenamefont{Tereshko et~al.}(1999)\citenamefont{Tereshko, Minasov,
  and Egli}}]{Tereshko:99b}
\bibinfo{author}{\bibfnamefont{V.}~\bibnamefont{Tereshko}},
  \bibinfo{author}{\bibfnamefont{G.}~\bibnamefont{Minasov}}, \bibnamefont{and}
  \bibinfo{author}{\bibfnamefont{M.}~\bibnamefont{Egli}}, \bibinfo{journal}{J.
  Am. Chem. Soc.} \textbf{\bibinfo{volume}{121}}, \bibinfo{pages}{3590}
  (\bibinfo{year}{1999}).

\bibitem[{\citenamefont{Sines et~al.}(2000)\citenamefont{Sines, McFail-Isom,
  Howerton, VanDerveer, and Williams}}]{Sines:00}
\bibinfo{author}{\bibfnamefont{C.~C.} \bibnamefont{Sines}},
  \bibinfo{author}{\bibfnamefont{L.}~\bibnamefont{McFail-Isom}},
  \bibinfo{author}{\bibfnamefont{S.~B.} \bibnamefont{Howerton}},
  \bibinfo{author}{\bibfnamefont{D.}~\bibnamefont{VanDerveer}},
  \bibnamefont{and} \bibinfo{author}{\bibfnamefont{L.~D.}
  \bibnamefont{Williams}}, \bibinfo{journal}{J. Am. Chem. Soc.}
  \textbf{\bibinfo{volume}{122}}, \bibinfo{pages}{11048}
  (\bibinfo{year}{2000}).

\bibitem[{\citenamefont{Hud et~al.}(1999)\citenamefont{Hud, {Sklen\'a\u r}, and
  Feigon}}]{Hud:99}
\bibinfo{author}{\bibfnamefont{N.~V.} \bibnamefont{Hud}},
  \bibinfo{author}{\bibfnamefont{V.}~\bibnamefont{{Sklen\'a\u r}}},
  \bibnamefont{and} \bibinfo{author}{\bibfnamefont{J.}~\bibnamefont{Feigon}},
  \bibinfo{journal}{J. Mol. Biol.} \textbf{\bibinfo{volume}{286}},
  \bibinfo{pages}{651} (\bibinfo{year}{1999}).

\bibitem[{\citenamefont{Denisov and Halle}(2000)}]{Denisov:00}
\bibinfo{author}{\bibfnamefont{V.~P.} \bibnamefont{Denisov}} \bibnamefont{and}
  \bibinfo{author}{\bibfnamefont{B.}~\bibnamefont{Halle}},
  \bibinfo{journal}{Proc. Natl. Acad. Sci. USA} \textbf{\bibinfo{volume}{97}},
  \bibinfo{pages}{629} (\bibinfo{year}{2000}).

\bibitem[{\citenamefont{Feig and Pettitt}(1999)}]{Feig:99}
\bibinfo{author}{\bibfnamefont{M.}~\bibnamefont{Feig}} \bibnamefont{and}
  \bibinfo{author}{\bibfnamefont{B.~M.} \bibnamefont{Pettitt}},
  \bibinfo{journal}{Biophys. J.} \textbf{\bibinfo{volume}{77}},
  \bibinfo{pages}{1769} (\bibinfo{year}{1999}).

\bibitem[{\citenamefont{Strahs and Schlick}(2000)}]{Strahs:00}
\bibinfo{author}{\bibfnamefont{D.}~\bibnamefont{Strahs}} \bibnamefont{and}
  \bibinfo{author}{\bibfnamefont{T.}~\bibnamefont{Schlick}},
  \bibinfo{journal}{J. Mol. Biol.} \textbf{\bibinfo{volume}{301}},
  \bibinfo{pages}{643} (\bibinfo{year}{2000}).

\bibitem[{\citenamefont{{\v S}tefl and Ko{\v c}a}(2000)}]{Stefl:00}
\bibinfo{author}{\bibfnamefont{R.}~\bibnamefont{{\v S}tefl}} \bibnamefont{and}
  \bibinfo{author}{\bibfnamefont{J.}~\bibnamefont{Ko{\v c}a}},
  \bibinfo{journal}{J. Am. Chem. Soc.} \textbf{\bibinfo{volume}{122}},
  \bibinfo{pages}{5025} (\bibinfo{year}{2000}).

\bibitem[{\citenamefont{Hamelberg et~al.}(2001)\citenamefont{Hamelberg,
  Williams, and Wilson}}]{Hamelberg:01}
\bibinfo{author}{\bibfnamefont{D.}~\bibnamefont{Hamelberg}},
  \bibinfo{author}{\bibfnamefont{L.~D.} \bibnamefont{Williams}},
  \bibnamefont{and} \bibinfo{author}{\bibfnamefont{W.~D.}
  \bibnamefont{Wilson}}, \bibinfo{journal}{J. Am. Chem. Soc.}
  \textbf{\bibinfo{volume}{32}}, \bibinfo{pages}{7745} (\bibinfo{year}{2001}).

\bibitem[{\citenamefont{McFail-Isom et~al.}(1999)\citenamefont{McFail-Isom,
  Sines, and Williams}}]{McFail-Isom:99}
\bibinfo{author}{\bibfnamefont{L.}~\bibnamefont{McFail-Isom}},
  \bibinfo{author}{\bibfnamefont{C.~C.} \bibnamefont{Sines}}, \bibnamefont{and}
  \bibinfo{author}{\bibfnamefont{L.~D.} \bibnamefont{Williams}},
  \bibinfo{journal}{Curr. Opin. Struct. Biol.} \textbf{\bibinfo{volume}{9}},
  \bibinfo{pages}{298} (\bibinfo{year}{1999}).

\bibitem[{\citenamefont{Chiu et~al.}(1999)\citenamefont{Chiu,
  Zaczor-Grzeskowiak, and Dickerson}}]{Chiu:99}
\bibinfo{author}{\bibfnamefont{T.~K.} \bibnamefont{Chiu}},
  \bibinfo{author}{\bibfnamefont{M.}~\bibnamefont{Zaczor-Grzeskowiak}},
  \bibnamefont{and} \bibinfo{author}{\bibfnamefont{R.~E.}
  \bibnamefont{Dickerson}}, \bibinfo{journal}{J. Mol. Biol.}
  \textbf{\bibinfo{volume}{292}}, \bibinfo{pages}{589} (\bibinfo{year}{1999}).

\bibitem[{\citenamefont{McConnell and Beverdige}(2000)}]{McConnell:00}
\bibinfo{author}{\bibfnamefont{K.~J.} \bibnamefont{McConnell}}
  \bibnamefont{and} \bibinfo{author}{\bibfnamefont{D.~L.}
  \bibnamefont{Beverdige}}, \bibinfo{journal}{J. Mol. Biol.}
  \textbf{\bibinfo{volume}{304}}, \bibinfo{pages}{803} (\bibinfo{year}{2000}).

\bibitem[{\citenamefont{Darden et~al.}(1993)\citenamefont{Darden, York, and
  Pedersen}}]{Darden:93}
\bibinfo{author}{\bibfnamefont{T.}~\bibnamefont{Darden}},
  \bibinfo{author}{\bibfnamefont{D.}~\bibnamefont{York}}, \bibnamefont{and}
  \bibinfo{author}{\bibfnamefont{L.}~\bibnamefont{Pedersen}},
  \bibinfo{journal}{J. Chem. Phys.} \textbf{\bibinfo{volume}{98}},
  \bibinfo{pages}{10089} (\bibinfo{year}{1993}).

\bibitem[{\citenamefont{Essmann et~al.}(1995)\citenamefont{Essmann, Perera,
  Berkowitz, Darden, Lee, and Pedersen}}]{Essmann:95}
\bibinfo{author}{\bibfnamefont{U.}~\bibnamefont{Essmann}},
  \bibinfo{author}{\bibfnamefont{L.}~\bibnamefont{Perera}},
  \bibinfo{author}{\bibfnamefont{M.~L.} \bibnamefont{Berkowitz}},
  \bibinfo{author}{\bibfnamefont{T.}~\bibnamefont{Darden}},
  \bibinfo{author}{\bibfnamefont{H.}~\bibnamefont{Lee}}, \bibnamefont{and}
  \bibinfo{author}{\bibfnamefont{L.~G.} \bibnamefont{Pedersen}},
  \bibinfo{journal}{J. Chem. Phys.} \textbf{\bibinfo{volume}{103}},
  \bibinfo{pages}{8577} (\bibinfo{year}{1995}).

\bibitem[{\citenamefont{Hockney and Eastwood}(1981)}]{Hockney:81}
\bibinfo{author}{\bibfnamefont{R.~W.} \bibnamefont{Hockney}} \bibnamefont{and}
  \bibinfo{author}{\bibfnamefont{J.~W.} \bibnamefont{Eastwood}},
  \emph{\bibinfo{title}{Computer Simulation Using Particles}}
  (\bibinfo{publisher}{McGraw-Hill}, \bibinfo{address}{New-York},
  \bibinfo{year}{1981}).

\bibitem[{\citenamefont{Wing et~al.}(1980)\citenamefont{Wing, Drew, Takano,
  Broka, Tanaka, Itakura, and Dickerson}}]{Wing:80}
\bibinfo{author}{\bibfnamefont{R.}~\bibnamefont{Wing}},
  \bibinfo{author}{\bibfnamefont{H.}~\bibnamefont{Drew}},
  \bibinfo{author}{\bibfnamefont{T.}~\bibnamefont{Takano}},
  \bibinfo{author}{\bibfnamefont{C.}~\bibnamefont{Broka}},
  \bibinfo{author}{\bibfnamefont{S.}~\bibnamefont{Tanaka}},
  \bibinfo{author}{\bibfnamefont{K.}~\bibnamefont{Itakura}}, \bibnamefont{and}
  \bibinfo{author}{\bibfnamefont{R.~E.} \bibnamefont{Dickerson}},
  \bibinfo{journal}{Nature} \textbf{\bibinfo{volume}{287}},
  \bibinfo{pages}{755} (\bibinfo{year}{1980}).

\bibitem[{\citenamefont{Arnott and Hukins}(1972)}]{Arnott:72}
\bibinfo{author}{\bibfnamefont{S.}~\bibnamefont{Arnott}} \bibnamefont{and}
  \bibinfo{author}{\bibfnamefont{D.~W.~L.} \bibnamefont{Hukins}},
  \bibinfo{journal}{Biochem. Biophys. Res. Communs.}
  \textbf{\bibinfo{volume}{47}}, \bibinfo{pages}{1504} (\bibinfo{year}{1972}).

\bibitem[{\citenamefont{Jorgensen et~al.}(1983)\citenamefont{Jorgensen,
  Chandreskhar, Madura, Impey, and Klein}}]{Jorgensen:83}
\bibinfo{author}{\bibfnamefont{W.~L.} \bibnamefont{Jorgensen}},
  \bibinfo{author}{\bibfnamefont{J.}~\bibnamefont{Chandreskhar}},
  \bibinfo{author}{\bibfnamefont{J.~D.} \bibnamefont{Madura}},
  \bibinfo{author}{\bibfnamefont{R.~W.} \bibnamefont{Impey}}, \bibnamefont{and}
  \bibinfo{author}{\bibfnamefont{M.~L.} \bibnamefont{Klein}},
  \bibinfo{journal}{J. Chem. Phys} \textbf{\bibinfo{volume}{79}},
  \bibinfo{pages}{926} (\bibinfo{year}{1983}).

\bibitem[{\citenamefont{Lavery and Sklenar}(1988)}]{Curves:}
\bibinfo{author}{\bibfnamefont{R.}~\bibnamefont{Lavery}} \bibnamefont{and}
  \bibinfo{author}{\bibfnamefont{H.}~\bibnamefont{Sklenar}},
  \bibinfo{journal}{J. Biomol. Struct. Dyn.} \textbf{\bibinfo{volume}{6}},
  \bibinfo{pages}{63} (\bibinfo{year}{1988}).

\bibitem[{\citenamefont{Martyna and Tuckerman}(1999)}]{Martyna:99}
\bibinfo{author}{\bibfnamefont{G.~J.} \bibnamefont{Martyna}} \bibnamefont{and}
  \bibinfo{author}{\bibfnamefont{M.~E.} \bibnamefont{Tuckerman}},
  \bibinfo{journal}{J. Chem. Phys.} \textbf{\bibinfo{volume}{110}},
  \bibinfo{pages}{2810} (\bibinfo{year}{1999}).

\bibitem[{\citenamefont{Cheatham et~al.}(1999)\citenamefont{Cheatham, Cieplak,
  and Kollman}}]{Cheatham:99}
\bibinfo{author}{\bibfnamefont{T.~E.} \bibnamefont{Cheatham},
  \bibfnamefont{III}},
  \bibinfo{author}{\bibfnamefont{P.}~\bibnamefont{Cieplak}}, \bibnamefont{and}
  \bibinfo{author}{\bibfnamefont{P.~A.} \bibnamefont{Kollman}},
  \bibinfo{journal}{J. Biomol. Struct. Dyn.} \textbf{\bibinfo{volume}{16}},
  \bibinfo{pages}{845} (\bibinfo{year}{1999}).

\bibitem[{\citenamefont{Cornell et~al.}(1995)\citenamefont{Cornell, Cieplak,
  Bayly, Gould, Merz, Ferguson, Spellmeyer, Fox, Caldwell, and
  Kollman}}]{Cornell:95}
\bibinfo{author}{\bibfnamefont{W.~D.} \bibnamefont{Cornell}},
  \bibinfo{author}{\bibfnamefont{P.}~\bibnamefont{Cieplak}},
  \bibinfo{author}{\bibfnamefont{C.~I.} \bibnamefont{Bayly}},
  \bibinfo{author}{\bibfnamefont{I.~R.} \bibnamefont{Gould}},
  \bibinfo{author}{\bibfnamefont{K.~M.} \bibnamefont{Merz}},
  \bibinfo{author}{\bibfnamefont{D.~M.} \bibnamefont{Ferguson}},
  \bibinfo{author}{\bibfnamefont{D.~C.} \bibnamefont{Spellmeyer}},
  \bibinfo{author}{\bibfnamefont{T.}~\bibnamefont{Fox}},
  \bibinfo{author}{\bibfnamefont{J.~W.} \bibnamefont{Caldwell}},
  \bibnamefont{and} \bibinfo{author}{\bibfnamefont{P.~A.}
  \bibnamefont{Kollman}}, \bibinfo{journal}{J. Am. Chem. Soc.}
  \textbf{\bibinfo{volume}{117}}, \bibinfo{pages}{5179} (\bibinfo{year}{1995}).

\bibitem[{\citenamefont{Mazur}(1997)}]{Mzjcc:97}
\bibinfo{author}{\bibfnamefont{A.~K.} \bibnamefont{Mazur}},
  \bibinfo{journal}{J. Comput. Chem.} \textbf{\bibinfo{volume}{18}},
  \bibinfo{pages}{1354} (\bibinfo{year}{1997}).

\bibitem[{\citenamefont{Mazur}(1999{\natexlab{a}})}]{Mzjchp:99}
\bibinfo{author}{\bibfnamefont{A.~K.} \bibnamefont{Mazur}},
  \bibinfo{journal}{J. Chem. Phys.} \textbf{\bibinfo{volume}{111}},
  \bibinfo{pages}{1407} (\bibinfo{year}{1999}{\natexlab{a}}).

\bibitem[{\citenamefont{Mazur}(2001{\natexlab{a}})}]{Mzbook:01}
\bibinfo{author}{\bibfnamefont{A.~K.} \bibnamefont{Mazur}}, in
  \emph{\bibinfo{booktitle}{Computational Biochemistry and Biophysics}}, edited
  by \bibinfo{editor}{\bibfnamefont{O.~M.} \bibnamefont{Becker}},
  \bibinfo{editor}{\bibfnamefont{A.~D.} \bibnamefont{MacKerell},
  \bibfnamefont{Jr}}, \bibinfo{editor}{\bibfnamefont{B.}~\bibnamefont{Roux}},
  \bibnamefont{and} \bibinfo{editor}{\bibfnamefont{M.}~\bibnamefont{Watanabe}}
  (\bibinfo{publisher}{Marcel Dekker}, \bibinfo{address}{New York},
  \bibinfo{year}{2001}{\natexlab{a}}), pp. \bibinfo{pages}{115--131}.

\bibitem[{\citenamefont{Mazur}(1999{\natexlab{b}})}]{Mzjmb:99}
\bibinfo{author}{\bibfnamefont{A.~K.} \bibnamefont{Mazur}},
  \bibinfo{journal}{J. Mol. Biol.} \textbf{\bibinfo{volume}{290}},
  \bibinfo{pages}{373} (\bibinfo{year}{1999}{\natexlab{b}}).

\bibitem[{\citenamefont{Anderson and Bauer}(1978)}]{Anderson:78}
\bibinfo{author}{\bibfnamefont{P.}~\bibnamefont{Anderson}} \bibnamefont{and}
  \bibinfo{author}{\bibfnamefont{W.}~\bibnamefont{Bauer}},
  \bibinfo{journal}{Biochemistry} \textbf{\bibinfo{volume}{17}},
  \bibinfo{pages}{594} (\bibinfo{year}{1978}).

\bibitem[{\citenamefont{Mazur}(1998)}]{Mzjacs:98}
\bibinfo{author}{\bibfnamefont{A.~K.} \bibnamefont{Mazur}},
  \bibinfo{journal}{J. Am. Chem. Soc.} \textbf{\bibinfo{volume}{120}},
  \bibinfo{pages}{10928} (\bibinfo{year}{1998}).

\bibitem[{\citenamefont{Mazur}(2000)}]{Mzjacs:00}
\bibinfo{author}{\bibfnamefont{A.~K.} \bibnamefont{Mazur}},
  \bibinfo{journal}{J. Am. Chem. Soc.} \textbf{\bibinfo{volume}{122}},
  \bibinfo{pages}{12778} (\bibinfo{year}{2000}).

\bibitem[{\citenamefont{Mazur}(2001{\natexlab{b}})}]{Mzjcc:01}
\bibinfo{author}{\bibfnamefont{A.~K.} \bibnamefont{Mazur}},
  \bibinfo{journal}{J. Comput. Chem.} \textbf{\bibinfo{volume}{22}},
  \bibinfo{pages}{457} (\bibinfo{year}{2001}{\natexlab{b}}).

\bibitem[{\citenamefont{Manning}(1978)}]{Manning:78}
\bibinfo{author}{\bibfnamefont{G.~S.} \bibnamefont{Manning}},
  \bibinfo{journal}{Q. Rev. Biophys.} \textbf{\bibinfo{volume}{2}},
  \bibinfo{pages}{179} (\bibinfo{year}{1978}).

\end{thebibliography}
\end{document}